\documentclass[aps,12pt,final,notitlepage,oneside,onecolumn,nobibnotes,nofootinbib,superscriptaddress,showpacs,centertags] {revtex4}
\usepackage{amsmath}
\usepackage{amsfonts}
\usepackage{graphicx}

\begin{document}
\title{RANDOM MATRIX ANALYSIS OF THE MONOPOLE STRENGTH DISTRIBUTION IN $^{208}$Pb}
\author{A. P. Severyukhin}
\affiliation{Bogoliubov Laboratory of Theoretical Physics,
             Joint Institute for Nuclear Research,
             141980 Dubna, Moscow region, Russia}
\affiliation{Dubna State University, Universitetskaya street 19,
             141980 Dubna, Moscow region, Russia}
\author{S. {\AA}berg}
\affiliation{Mathematical Physics, LTH, Lund University,
PO Box 118, S-22100 Lund, Sweden}
\author{N. N. Arsenyev}
\affiliation{Bogoliubov Laboratory of Theoretical Physics,
             Joint Institute for Nuclear Research,
             141980 Dubna, Moscow region, Russia}
\author{R. G. Nazmitdinov}
\affiliation{Departament de F\'{\i}sica, Universitat de les Illes Balears,
E-07122 Palma de Mallorca, Spain}
\affiliation{Bogoliubov Laboratory of Theoretical Physics,
             Joint Institute for Nuclear Research,
             141980 Dubna, Moscow region, Russia}
\affiliation{Dubna State University, Universitetskaya street 19,
             141980 Dubna, Moscow region, Russia}
\author{K. N. Pichugin}
\affiliation{Kirensky Institute of Physics,
             660036 Krasnoyarsk, Russia}
\begin{abstract}
We study  statistical properties of the $0^+$ spectrum of
$^{208}$Pb in the energy region $E_x\leq20$ MeV. We use the Skyrme
interaction SLy4 as our model Hamiltonian to create a
single-particle spectrum and to analyze excited states. The
finite-rank separable approximation for the particle-hole
interaction enables us to perform the calculations in large
configuration spaces. We show that while the position of the
monopole resonance centroid is determined by one phonon
excitations of $0^+$, the phonon-phonon coupling is crucial for
the description of a strength distribution of  the $0^+$ spectrum.
In fact, this coupling has an impact on the spectral rigidity
$\Delta_3(L)$ which is shifted towards the random matrix limit of
the Gaussian orthogonal ensembles.
\end{abstract}


\maketitle
\section{Introduction}
Nuclear Giant Resonances (GR) are the subject of numerous
investigations over several decades \cite{dstadta}. Some of the
basic features such as centroids and collectivity (in terms of the
sum rules) are reasonably well understood within microscopic
theories \cite{BM,M,S}.  As yet we have no answer on the question
how a collective mode like the GR dissipates its energy.

According to accepted wisdom, GRs are essentially excited by an
external field through a one-body interaction. It is, therefore,
natural to describe these states as collective \rm{1p-1h} states.
Once excited, the GR progresses to a fully equilibrated system via
direct particle emission and by coupling to more complicated
configurations (\rm{2p-2h}, \rm{3p-3h}, etc). The former mechanism
gives rise to an escape width. It is expected that the decay
evolution along the hierarchy of more complex configurations till
compound states determines spreading widths. A full description of
this decay represents a fundamental problem which is, however,
difficult to solve (if even  is possible at all ?) due to
existence of many degrees of freedom for many-body quantum system
such as a nucleus. Therefore, to  gain an insight into the nature
of  GR spreading we have to introduce various approximations or a
model, which validity depends on a primal feasibility.

In general, the description of GR properties is based on the
analysis of the electromagnetic strength distribution in an energy
interval which is large enough to catch hold of basic GR features
that is under investigation. An obvious requirement to the model
consideration is to use configurations with various degree of
complexity. Evidently, the more complex configuration is
considered, the more cumbersome task should be solved. Therefore,
a natural question arises: what type of  a particular complex
configuration should be enough in order to understand a gross
structure of a particular GR which data are available  in modern
experiments? In addition, once this complex configuration is
defined one can further ask about statistical properties of states
that compose a GR strength distribution. As a result, one could
illuminate the role of various correlations that diminish  an
importance of a specific excitation that determines a centroid
position of a specific GR.

To answer these questions we will employ two approaches. On the
one hand, the random matrix theory (RMT)~\cite{B81,M91,GMW,HW,Gom}
provides necessary tools to shed light on the spectral properties
and the distribution of transition-strength properties, when
specific features become not of a primary importance.
The RMT assumes only that the nuclear Hamiltonian belongs to an ensemble
of random matrices that are consistent with the fundamental
symmetries of the system. In particular,
in the case of the time-reversal symmetry, the relevant ensemble
is the Gaussian orthogonal ensemble (GOE).
On the other hand, to understand the
fragmentation of high-lying states it is necessary also to exploit
nuclear structure models that are based on the microscopic
many-body theory, where the effects of the residual interaction on
the statistics must be studied in large model spaces. Introducing
a residual interaction in general implies a transition to
GOE-properties above some excitation energy~\cite{A}.

The quasiparticle-phonon model (QPM)~\cite{S} offers an attractive
framework for such studies. The separable form of the residual
interaction of a model Hamiltonian allows to diagonalize it in a
space spanned by states composed of one, two and three phonons
considered in the random phase approximation (RPA). We would like
to mention here the RMT analysis of statistical properties of a
pygmy dipole resonance within the QPM, based on the Woods-Saxon
potential \cite{VP}. It is desirable, however, to use a unified
approach in which a mean field and a residual interaction are
treated at the same footing in order to avoid any artifacts
\cite{sfkh78,BNY}. For our  purposes we choose the modern
development of the QPM, a finite rank separable approximation
(FRSA)~\cite{gsv98,ssvg02,svg08}. The FRSA follows the basic QPM
ideas, but the single-particle (sp) spectrum and the residual
interaction are calculated with the Skyrme forces. This approach
enables us to consider
 a coupling between the one- and two-phonon components of the wave
functions~\cite{svg04}.  It was successfully used to study the
properties of the low-lying states and giant resonances within the
RPA and beyond~\cite{gsv98,ssvg02,svg08,svg04,sap12,sapw14}.

By means of this approach and by the RMT tools
we attempt in this paper to understand the complex structure
observed in the $0^{+}$ spectrum of the
doubly magic nucleus $^{208}$Pb in the region of  the isoscalar
giant monopole resonance (ISGMR). This strength distribution is
extensively studied in many experiments~\cite{Y99,Y04,P13,P14}.
The experimental properties have been described within the RPA
with the Skyrme interactions (for a review see,  for example, ~Ref.\cite{rev2007}).
In this system the effect of the anharmonicity is expected to be small.
Contrary to the expectations, we will show
the importance of the phonon-phonon coupling (PPC) effects for the
statistical properties of the spectrum calculated with the aid of
the Skyrme SLy4 interaction, taken as an example.

\section{The Model}
For the analysis of the doubly magic nucleus we impose a spherical
symmetry on the sp wave functions in our HF
calculations. The continuous part of the sp spectrum is
discretized by diagonalizing the HF Hamiltonian on a harmonic
oscillator basis. The cut-off of the continuous part is at the
energy of 100~MeV. As the parameter set, we use the Skyrme force
SLy4~\cite{sly} which was adjusted to reproduce the nuclear matter
properties, as well as nuclear charge radii, binding energies of
doubly magic nuclei. The residual particle-hole interaction is
obtained as the second derivative of the energy density functional
with respect to the particle density. By means of the standard
procedure~\cite{tebdns05} we obtain the familiar RPA equations in
the \rm{1p-1h} configuration space. The eigenvalues of the RPA
equations are found numerically as the roots of a relatively
simple secular equation within the FRSA ~\cite{gsv98}. Since the
FRSA enables to us to use the large \rm{1p-1h} space, there is not
need in effective charges.

Using the basic QPM ideas in the simplest case of the
configuration mixing~\cite{S}, we construct the wave functions
from a linear combination of one-phonon and two-phonon
configurations states as
\begin{eqnarray}
\label{wf} \Psi_\nu(JM)=\biggl\{\sum\limits_i R_i(J\nu)Q_{JMi}^{+}
+\sum\limits_{\lambda_1i_1\lambda_2i_2}P_{\lambda_2i_2}^{\lambda_1i_1}(J\nu)
&\left[Q_{\lambda_1\mu_1 i_1}^{+}Q_{\lambda_2\mu_2
i_2}^{+}\right]_{JM}\biggr\}\mid 0\rangle,
\end{eqnarray}
where $Q_{\lambda\mu i}^{+} |0\rangle$ is the RPA excitation
having energy $\omega_{\lambda i}$; $\lambda$ denotes the total
angular momentum and $\mu$ is its z-projection in the laboratory
system. The ground state is the RPA phonon vacuum $\mid0\rangle $.
The normalization condition for the wave functions~(\ref{wf}) yields
\begin{equation}
\label{norma2ph} \sum\limits_iR_i^2(J\nu)+ 2\sum_{\lambda _1i_1
\lambda _2i_2} (P_{\lambda _2i_2}^{\lambda _1i_1}(J\nu))^2=1.
\end{equation}
The variational principle leads to a set of linear equations for
the unknown amplitudes $R_i(J\nu)$ and
$P_{\lambda_2i_2}^{\lambda_1i_1}(J\nu)$~\cite{svg04}:
\begin{equation}
 \label{2pheq1}
(\omega _{Ji}-E_\nu )R_i(J\nu )+\sum_{\lambda _1i_1 \lambda_2i_2}
U_{\lambda _2i_2}^{\lambda _1i_1}(Ji) P_{\lambda_2i_2}^{\lambda
_1i_1}(J\nu )=0,
\end{equation}
\begin{equation}
 \label{2pheq2}
 \sum\limits_iU_{\lambda _2i_2}^{\lambda _1i_1}(Ji)R_i(J\nu ) +
2(\omega _{\lambda _1i_1}+\omega _{\lambda _2i_2}-E_\nu )
P_{\lambda _2i_2}^{\lambda _1i_1}(J\nu)=0.
\end{equation}
The rank of the set of linear equations is equal to the number of
one- and two-phonon configurations included in the wave function
(\ref{wf}). To resolve this set it is required to compute the
coupling matrix elements
\begin{equation}
U_{\lambda _2i_2}^{\lambda _1i_1}(J i) = \langle 0| Q_{J i } H
\left[ Q_{\lambda _1i_1}^{+}Q_{\lambda _2i_2}^{+}\right] _{J} |0
\rangle
\end{equation}
between one- and two-phonon configurations
(see details in Ref.~\cite{svg04}). Evidently, the nonzero matrix elements
$U_{\lambda _2i_2}^{\lambda _1i_1}(J i)$ result in the inclusion
of the PPC effects. Eqs.(\ref{2pheq1}) and
(\ref{2pheq2}) have the same form as the QPM equations~\cite{S}.
It is important to stress, however, that
the sp spectrum and the parameters of the
residual \rm{p-h} interaction are calculated with the chosen
Skyrme forces, without any further adjustments.

The excitation operator of the ISGMR is defined as
\begin{eqnarray}
\hat{M}_{L=0}=\sum\limits_{i=1}^{A}r^2_i.
\end{eqnarray}
The wave functions~(\ref{wf}) allow us to determine the transition
probabilities
$\left|\langle0^{+}_{\nu}|\hat{M}_{L=0}|0^{+}_{g.s.}\rangle\right|^2$.
The matrix elements for direct excitation of two-phonon components from
the ground state are about two orders of magnitude smaller as compared
to the excitation of one-phonon components~\cite{S}. Therefore, they
are neglected in our calculation of the transition probabilities.
The RPA analysis of the ISGMR shows that 96~\%
 of the non energy-weighted sum rules~(NEWSR) is
located in the energy region
$E_{x}=10.5-18$~MeV. To build the wave functions~(\ref{wf})
of the excited $0^+$ states up to 20~MeV we take into account all
one- and two-phonon configurations
$[\lambda^{\pi}_{i_1}\otimes\lambda^{\pi}_{i_2}]_{RPA}$
that are constructed from the
$0^+$, $1^-$, $2^+$, $3^-$ and  $4^+$ phonons with energies below
25~MeV for computational convenience. The high-energy
configurations plays a minor role in our calculations. It is
noteworthy that the pair-transfer mode (see, e.g.,
Refs.~\cite{J87,H15}) is outside the present work.

Properties of the low-energy two-phonon $0^+$ states are reflected
in the deviation from the harmonic picture for the multiphonon
excitations~\cite{L97,P99}. It is interesting to study the
energies, reduced transition probabilities of
the $[2_1^+]_{RPA}$, $[3_1^-]_{RPA}$ and $[4_1^+]_{RPA}$ states
which are the important ingredients of our
calculations of the two-phonon $0^+$ states of $^{208}$Pb. The
results obtained within the one-phonon approximation are compared with the
experimental data~\cite{H82,S83} in Table~1. There is
a satisfactory description of the reduced transition
probabilities.  The overestimate of the experimental energies
indicates on some missing mechanisms. In our consideration
we consider the PPC as the one that might improve the description.

The strength distribution of ISGMR is displayed in Fig.1. Both,
experimental~\cite{P13} and theoretical results show the
fragmentation and splitting of the ISGMR strength. The coupling
between the one- and two-phonon states yields a noticeable
redistribution of the ISGMR strength in comparision with the RPA
results. In particular, the coupling decreases the NEWSR till
78~\%   in the ISGMR region ($E_{x}=10.5-18$~MeV). Also, the PPC
induces  the 1~MeV downward shift of the main peak. There are the
low-energy part, the main peak and the high-energy tail. The
coupling produces a shift of order 11~\% (7~\%) of the NEWSR from
the ISGMR region to the high (lower) energy region. The strength
distribution of the ISGMR obtained within the PPC is rather close
to the experimental distribution~\cite{P13}. Our analysis shows
that the major contribution to the strength distribution is
brought about by the coupling between the $[0^+]_{RPA}$ and
$[3^-\otimes3^-]_{RPA}$ components. We recall that the importance
of the complex configurations for the interpretation of basic
peculiarities of the ISGMR strength distribution of $^{208}$Pb was
already qualitatively discussed in the framework of simple model
\cite{GB79,BB81}. Our calculations give the same tendency.

We turn now to the mechanism that dominates in  the low-energy
part of the $0^+$ spectrum. There is no $[0^+]_{RPA}$ state below
10.2~MeV. The extension of the variational space from the standard
RPA to  two-phonon configurations result in a formation of the
low-lying $0^+$ states. The $[3_1^-]_{RPA}$ state is the lowest
excitation which leads to the minimal two-phonon energies and the
maximal matrix elements coupling between one- and two-phonon
configurations. Since the PPC induces a downward shift of the
$0^+_{1}$ energy, the energy state at 6.5~MeV is very close to the
value $\sim 2 \hbar\Omega$ with $\hbar\Omega=E_{3_1^-}^{RPA}$ (see
Table 1). Our analysis suggests the dominance ($\geq$85\%) of the
octupole
 $[3_1^-\otimes3_1^-]_{RPA}$,
$[3_1^-\otimes3_2^-]_{RPA}$ and $[3_1^-\otimes3_3^-]_{RPA}$
configurations in the wave functions of the excited
$0_{1}^+$, $0_{2}^+$ and $0_{3}^+$ states, respectively.
The collective character of the $0_{1}^+$ state is mainly due to
their coupling to the ISGMR, produced by the
$[0_{4}^+]_{RPA}$ state. In particular, the wave function normalization
of the $0_{1}^+$ state contains 4{\%} of the $[0_4^+]_{RPA}$. This small
change in structure has a large impact on the
$\left|\langle0^{+}_{1}|\hat{M}_{L=0}|0^{+}_{g.s.}\rangle\right|^2$ value,
see Fig.1. The lowest two-phonon $0^+$ state was first observed as
the lowest-spin member of the $[3_1^-\otimes3_1^-]$ multiplet in
Ref.~\cite{Y96}. This fact was confirmed by the QPM analysis ~\cite{P99}.
%
%
\section{Spectral statistics}
Let us study statistical properties of
the $0^+$ spectrum up to 20 MeV. We examine the spectra calculated with and
without the PPC effects, i.e., the cases of $U\neq0$ and $U=0$, respectively.
 Fig.~2 displays the PPC impact on the $0^+$
energies. As was mentioned above, the coupling shifts down the part of $0^+$ states
 and modifies the level density. To elucidate the role of the residual interaction we also
 consider the $0^+$ spectrum of unperturbed \rm{1p-1h} and \rm{2p-2h} states  (see Fig.~2a).
 Note that the level density of
the unperturbed \rm{3p-3h} states is much smaller than the \rm{2p-2h} ones.
As can be seen from Fig.~2, the difference between the unperturbed \rm{p-h}
and the $U=0$ spectra is remarkable. The downward shift of the $U=0$ spectrum
is due to the residual interaction in the RPA framework. The coupling  does not
lead to visible spectrum changes. However, it brings important correlations that
affect the spectral statistics.

The three spectra are analyzed within the RMT
that enables to us to
study the statistical laws governing fluctuations
that, in general, can have very different origins. Starting from the
spectrum $E_i$, one can construct the staircase function
$N(E)$ which is defined as the state number below the energy $E$.
The function $N(E)$ can be separated in a smooth part $S(E)$ and
the fluctuating part $N_{\rm fluct}(E)$, where the integral
of $N_{\rm fluct}(E)$ is zero. The function $S(E)$  can be
determined either from semiclassical arguments or using a
polynomial for $N(E)$. To get a constant mean spacing of levels,
we employ the unfolded spectrum defined by the mapping $x_i=S(E_i)$.
Note that the values $s_i=x_{i+1}-x_i$ are introduced as the spacings.
We use two typical measures to analyze the fluctuation properties
of unfolded spectrum: the nearest-neighbor spacing distribution (NNSD)
and the spectral rigidity of Dyson and Metha, the $\Delta_{3}$
statistics~\cite{DM63}.

Due to the unfolding we have
\begin{equation}
\int_0^{\infty}sP(s)ds=1.
\end{equation}
If the unfolded energies $x_i$ are in a regular system then the NNSD is known as
the Poisson distribution,
\begin{equation}
P(s)=e^{-s}.
\end{equation}
In the GOE, i.e. the energies are in a chaotic system, the NNSD is
approximately  given as the Wigner distribution~\cite{M91},
\begin{equation}
P(s)=(\pi/2)s\exp(-\pi{s^2}/4).
\end{equation}
As can be seen from Fig.~3, for the unperturbed \rm{p-h} spectrum
we obtain a behavior close to the Poisson distribution, expected
for uncorrelated energies. For the case of $U\neq0$,  the
statistics is changing to the GOE limit. This fact indicates the
onset of correlations that redistribute a $[0^+]_{RPA}$ strength
over two-phonon components constructed by phonons with the other
multipolarities. Indeed, the comparison of the NNSD without and
with the coupling illuminates this fact evidently. At $U=0$ the
spectrum is characterized by the Poisson (uncorrelated)
statistics. The coupling between the one- and two-phonon
components modifies the spectrum, and the NNSD becomes close to
the Wigner surmise.

Another measure of  correlations is the $\Delta_{3}$ statistics defined as
\begin{equation}
\label{delta3}
\Delta_{3}(\alpha,L)=\min_{A,B}\frac{1}{L}\int\limits_{\alpha}^{\alpha+L}
\left[N(x)-(Ax+B)\right]^2dx.
\end{equation}
It characterizes the deviation of the staircase function $N(x)$
from a straight line, and a rigid unfolded spectrum corresponds to
smaller values of $\Delta_{3}$, while a soft spectrum has a larger
$\Delta_{3}$. In fact, for a given L, smaller values of
$\Delta_{3}$ imply stronger long-range correlations between the
levels.

For the sake of convenience,
the function $\Delta_{3}(\alpha,L)$, averaged over $n_{\alpha}$
intervals $(\alpha,\alpha+L)$
\begin{equation}
\bar{\Delta}_{3}(L)=\frac{1}{n_{\alpha}}\sum\limits_{\alpha}\Delta_{3}(\alpha,L) ,
\end{equation}
can be easily calculated from the number statistics, $n(L)$, which
is the number of levels in an energy interval of length $L$
\begin{equation}
\bar{\Delta}_{3}(L)=\frac{2}{L^4}\int\limits_{0}^{L}
\left(L^{3}-2L^{2}r+r^{3}\right)\Sigma^{2}(r)dr,
\end{equation}
\begin{equation}
\Sigma^{2}(L)=\left\langle[n(L)-\langle
n(L)\rangle]^2\right\rangle.
\end{equation}
For an uncorrelated spectrum one has
\begin{equation}
\bar{\Delta}_{3}(L)=L/15,
\end{equation}
while for  the GOE it is
\begin{equation}
\bar{\Delta}_{3}(L)\approx \frac{1}{\pi^2}(\ln L -0.0687)
\end{equation}
in the $L\gg1$ limit. Fig.4 demonstrates the evolution of the
$\Delta_{3}$ measure from the uncorrelated states to the GOE
limit,  when the PPC effects are only responsible for the
statistical correlations. These correlations dissolve the
collective ISGMR in the sea of the fragmented two-phonon $0^+$
components created by the other multipolarities.

\section{Summary}
With the aid of a finite rank separable approximation we have
analyzed the strength distribution of $0^+$ states ($E_x\leq20$
MeV) of $^{208}$Pb. To simulate the mean-field we have used the
SLy4 Skyrme interaction. To analyse the $0^+$ excitations we take
into account all RPA states with $\lambda^{\pi}$ =$0^+$ , $1^-$,
$2^+$, $3^-$, $4^+$. Within the RPA approach the centroid location
of the ISGMR is found at $E\sim14.7$ MeV. On the other hand, we
have demonstrated that the coupling between one- and two-phonon
terms in the wave functions of excited states is crucially
important for the interpretation of the strength distribution of
the ISGMR in the energy interval $E_x\approx 10.5-18$ MeV. The
results of the calculated transition-strength distribution are
generally in a reasonable agreement with the experimental data.

The RMT measures such as the NNSD and the $\Delta_{3}$ function
indicate a transition towards GOE as soon as the coupling is
switched on. It appears that the presence of two-phonon components
in our wave function, in addition to the one-phonon ones, already
enables to us to describe the gross strength distribution of the
ISGMR in the experimentally available energy interval. We observed
that the major contribution that evolves the system under
consideration to the GOE limit is brought about by two-pnonon
components of the octupole nature. In fact, their number exceeds
essentially the numbers of two-phonon components that are
constructed from phonons with the other multipolarities. A further
systematic statistical studies of the impact of the phonon-phonon
coupling on the vibrational spectra and the $E\lambda$-transition
strengths is clearly necessary and is in progress.
%
%
\section*{Acknowledgments}
A.P.S. thanks for the hospitality the Division of Mathematical
Physics, Lund University, where a part of this work has been done.
This work was partly supported by Russian Foundation for Basic
Research under Grant nos. 16-52-150003 and 16-02-00228.
%
%

%
\newpage
%
\begin{table}[]
\caption[]{Energy and $B(E\lambda)$ values for up-transitions to
the $\lambda_{1}^{\pi}$ states in $^{208}$Pb. Experimental data
are taken from Refs.~\cite{H82,S83}.}
\begin{tabular}{ccccc}
\hline\noalign{\smallskip}
 $\lambda_{1}^{\pi}$ &\multicolumn{2}{c}{Energy, MeV}& \multicolumn{2}{c}{$B(E\lambda;0^+_{g.s.}\rightarrow\lambda_{1}^{\pi})$, e$^2$b$^{\lambda}$} \\
                       &    Exp.       &  RPA       &     Exp.       &  RPA                      \\
\hline\noalign{\smallskip}
 $3^{-}_{1}$           &    2.62       &   3.6      & $0.611\pm0.012$ & 0.93\\
 $2^{+}_{1}$           &    4.09       &   5.2      & $0.318\pm0.016$ & 0.34\\
 $4^{+}_{1}$           &    4.32       &   5.6      & $0.155\pm0.011$ & 0.15\\
\hline\noalign{\smallskip}
\end{tabular}
\end{table}

\newpage

\begin{figure}[t!]
\includegraphics[width=1.0\columnwidth]{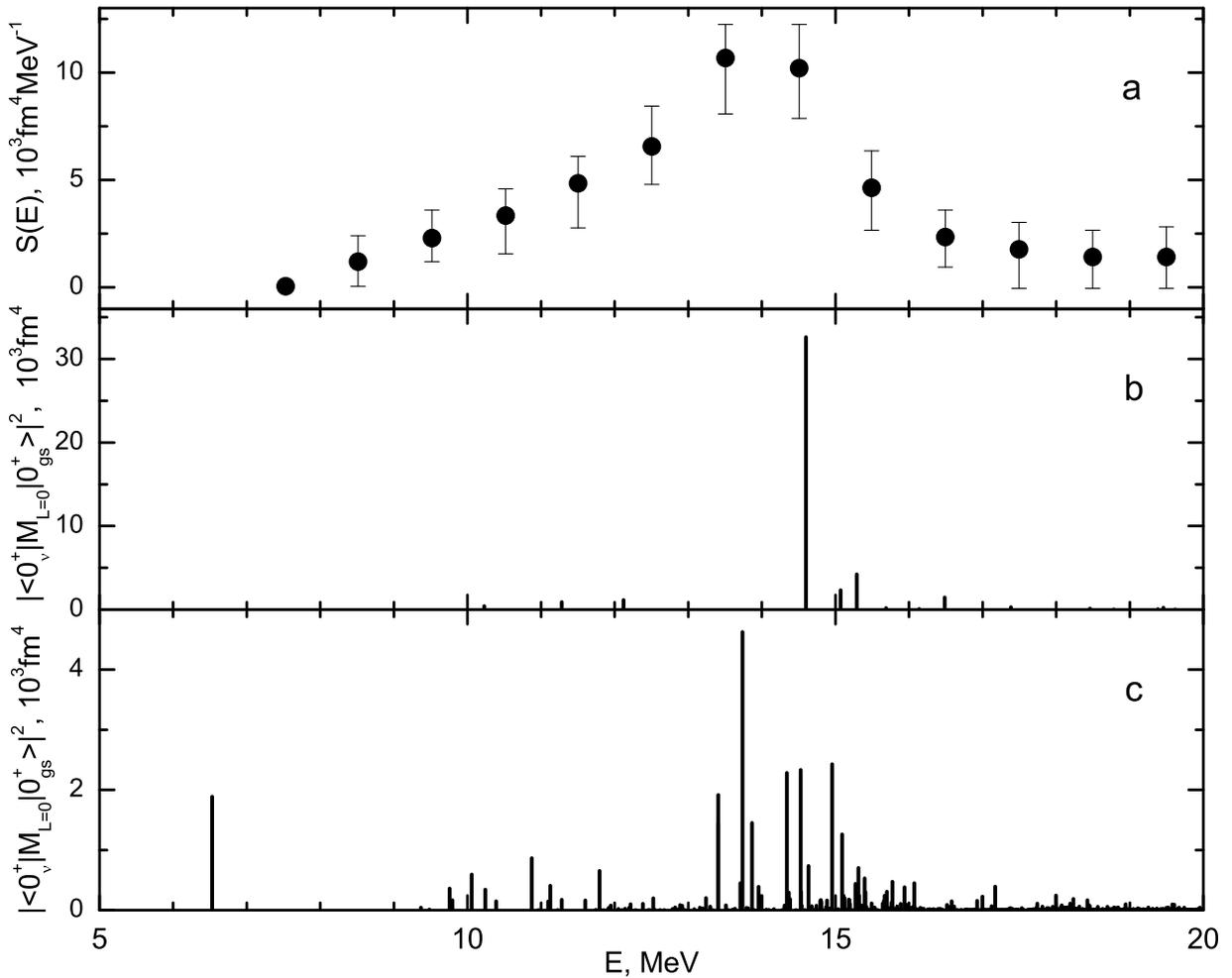}
\caption{The PPC effect on the isoscalar
monopole strength distribution in $^{208}$Pb.
Panel (a): experimental strength distribution is
taken from Ref.~\cite{P13}.
Panels (b) and (c) correspond to the calculations
within the RPA and taking into account
the PPC, respectively.}
\end{figure}

\newpage
%
\begin{figure*}[t!]
\includegraphics[width=1.0\columnwidth]{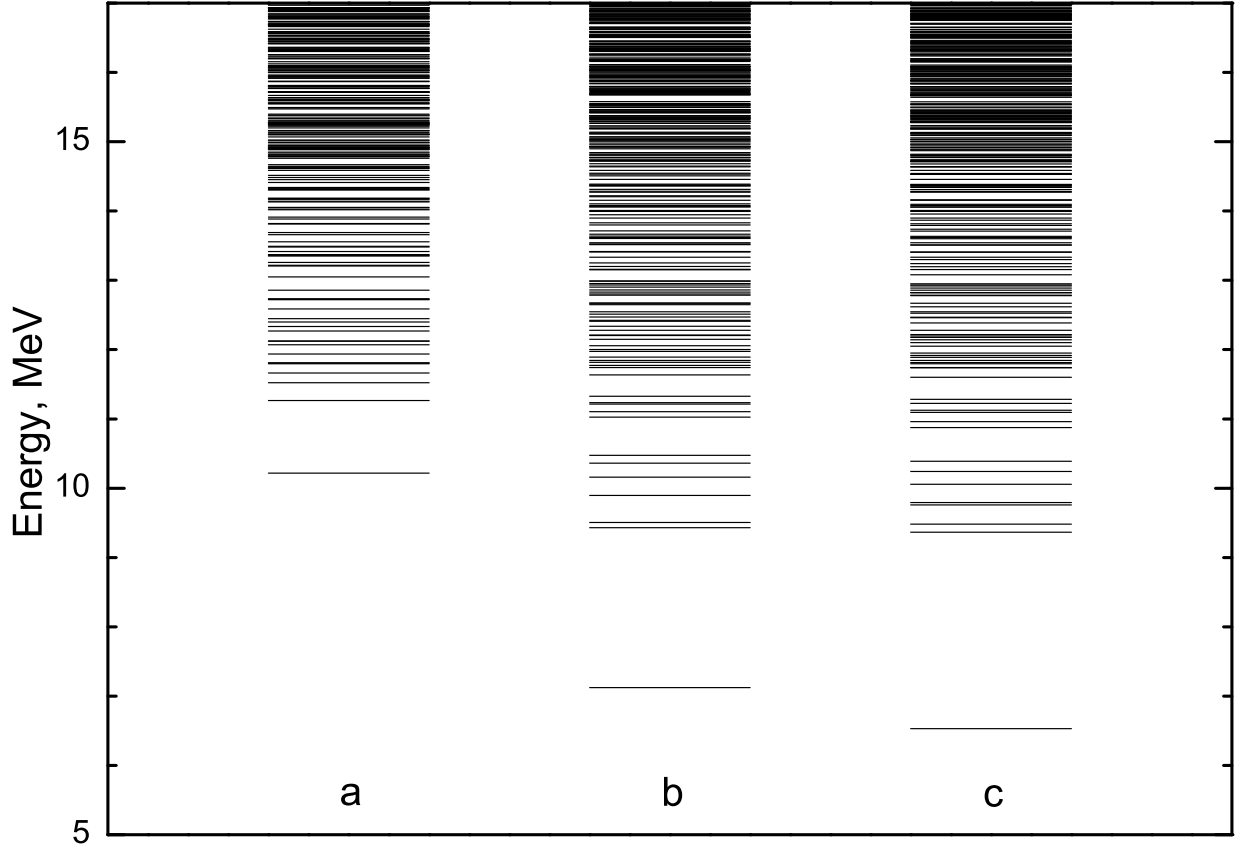}
\caption{Calculated spectra of the $0^+$ states of $^{208}$Pb.
The unperturbed \rm{1p-1h} and \rm{2p-2h} energies
are shown in column (a). Columns (b) and (c) correspond
to the calculations without and with the effects of
the PPC, respectively.}
\end{figure*}

\newpage
%
\begin{figure*}[t!]
\includegraphics[width=0.8\columnwidth]{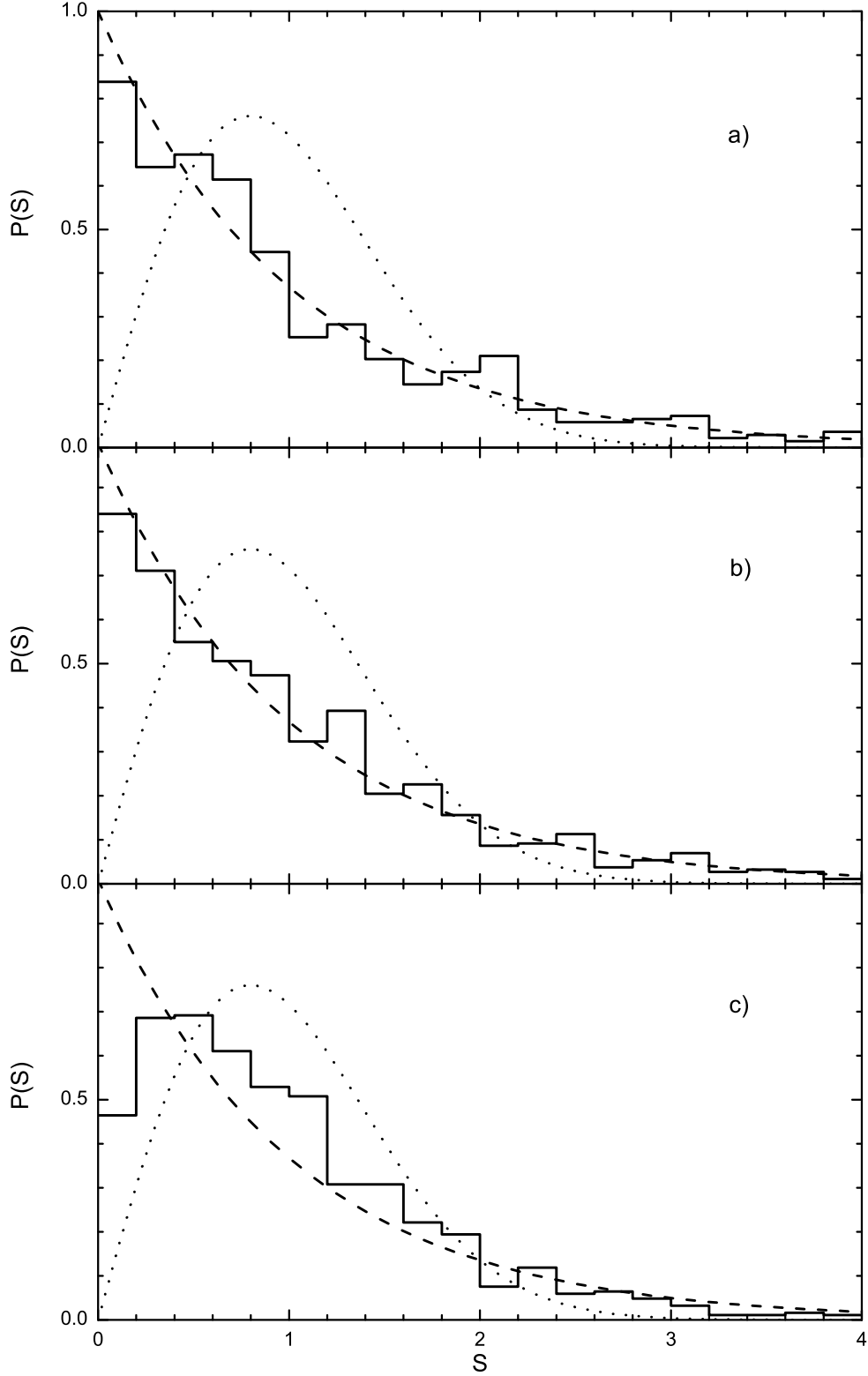}
\caption{The nearest-neighbor spacing
distribution for calculated spectra of the $0^+$ states
of $^{208}$Pb. The case of unperturbed \rm{1p-1h}
and \rm{2p-2h} energies is shown in panel (a).
Panels (b) and (c) correspond to the calculations without
and with the PPC effects, respectively.
The dotted line is the GOE limit and the dashed line is
the Poisson statistic.}
\end{figure*}

\newpage
%
\begin{figure*}[t!]
\includegraphics[width=0.8\columnwidth]{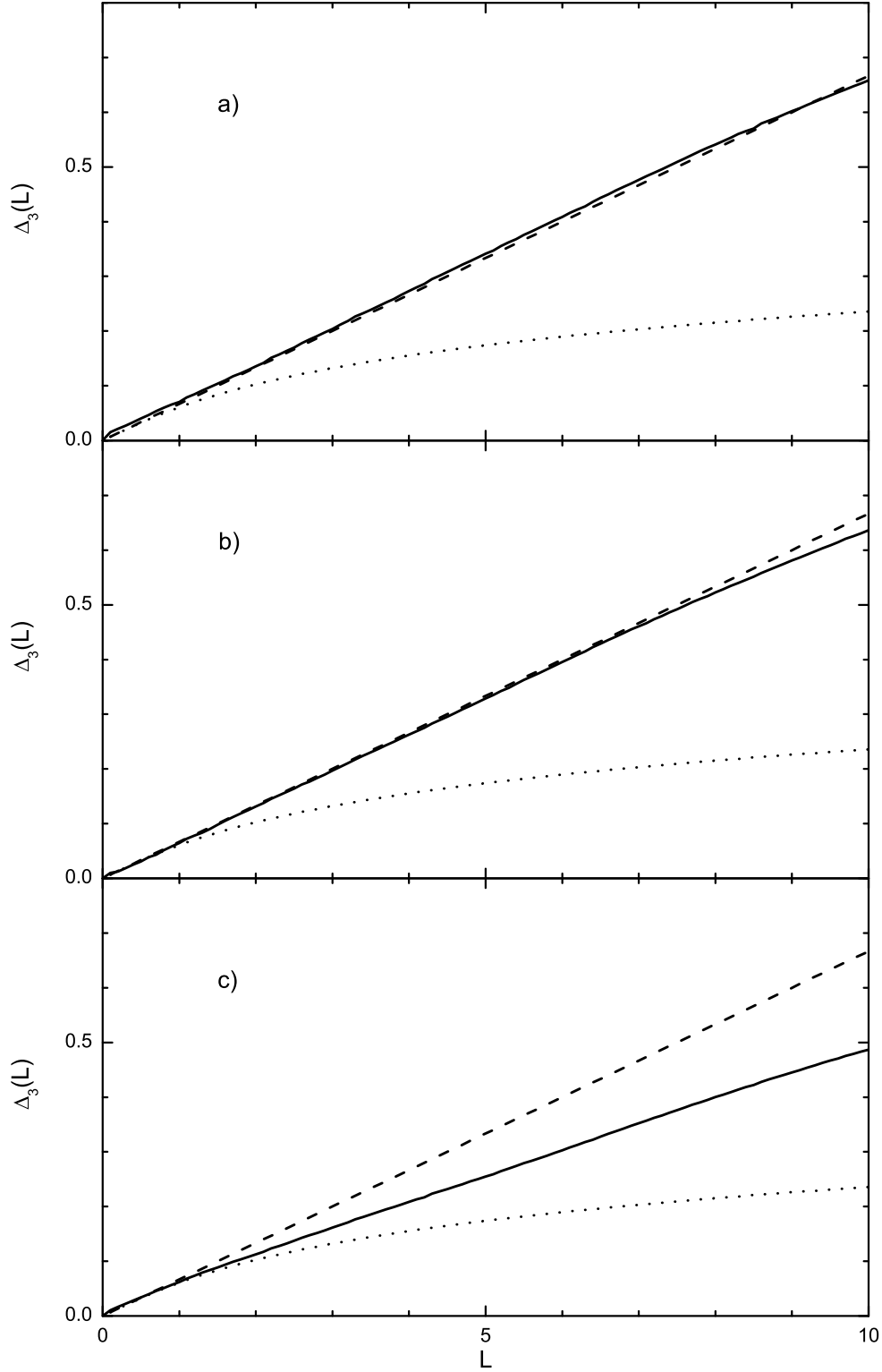}
\caption{The spectral rigidity $\Delta_3(L)$ for calculated spectra
of the $0^+$ states of $^{208}$Pb.
The case of unperturbed \rm{1p-1h}
and \rm{2p-2h} energies is shown in panel (a).
Panels (b) and (c) correspond to the calculations without
and with the PPC effects, respectively.
The dotted line is
the GOE limit and the dashed line is the Poisson statistic.}
\end{figure*}
\end{document}